\documentclass[aps,prb,twocolumn,floatfix]{revtex4}
\usepackage[final]{graphicx}
\usepackage{bm}
\usepackage{float}
\usepackage{afterpage}

\begin{document}

\title{Topological phase in one-dimensional interacting fermion system}
\author{Huaiming Guo$^{1,2}$ and Shun-Qing Shen$^2$}
\affiliation{$^1$Department of Physics, Beihang University, Beijing, 100191, China}
\affiliation{$^2$Department of Physics, The University of Hong Kong, Pokfulam Road, Hong
Kong}

\begin{abstract}
We study a one-dimensional interacting topological model by means of exact
diagonalization method. The topological properties are firstly examined with
the existence of the edge states at half-filling. We find that the
topological phases are not only robust to small repulsive interactions but
also are stabilized by small attractive interactions, and also finite
repulsive interaction can drive a topological non-trivial phase into a
trivial one while the attractive interaction can drive a trivial phase into
a non-trivial one. Next we calculate the Berry phase and parity of the bulk
system and find that they are equivalent in characterizing the topological
phases. With them we obtain the critical interaction strengths and
construct part of the phase diagram in the parameters space. Finally we
discuss the effective Hamiltonian at large-$U$ limit and provide additional
understanding of the numerical results. Our these results could be realized
experimentally using cold atoms trapped in the 1D optical lattice.
\end{abstract}

\maketitle


\section{Introduction}

The finding of time-reversal invariant topological insulators (TIs) has
become an exciting event in condensed matter physics. Since then lots of
works have been carried out theoretically and experimentally, predicting and
verifying many exotic physical properties exhibited by TIs \cite%
{moore1,hasan1,xlqi1}. The key feature of TIs is the existence of the robust
edge states determined by the bulk topological property, which can be
described by $Z_{2}-$ valued topological invariants\cite%
{kane1,kane2,moore2,kane3}. The original definition of TIs is for
non-interacting band structures and the relevant physics has been well
understood. Thereafter one of the subjects that need to be explored further
is the effects of electron correlations on TIs.

In the situation where many-body interactions exist, the definition of TIs
from the topological field theory, which is the presence or absence of a
topological term in the effective electromagnetic action, is generally valid%
\cite{xlqi2}. In addition, the method using the Green's functions to
construct the topological invariants is also appliable\cite%
{volovik,gurarie,zwang}. Yet the main difficulty is still to deal with the
interactions properly. At the mean-field level, it has been shown that
interactions can change the trivial insulators into non-trivial ones \cite%
{raghu,jwen,cweeks}. It is also been proposed that TIs and new topological
phases may appear in the systems with considerable interactions such as $4d$
or $5d$ transition metal oxides \cite{pesin,shitade}. Recently several
numerical simulations and analytical works were performed on the interacting two-dimensional
Kane-Mele model and studied the interplay of spin-orbit coupling and Coulomb
repulsion \cite{varney,zymeng,hohenadler,jxli,dzheng,yyamaji,dhlee,rachel1,rachel2}. These
studies show that the Hubbard repulsive interaction can transform the TI of
Kane-Mele model to either the spin liquid phase or antiferromagnetic
insulating phase depending on the strength of the spin-orbit coupling. A
study using the Lanczos algorithm concludes that the topological properties
have already manifested themselves in small systems and therefore can be
studied numerically via exact diagonalization (ED) and observed
experimentally\cite{varney}.

There also appear works addressing the question that how the presence of
interactions changes the classification of the topological phases. For
noninteracting systems, five symmetry classes are topological nontrivial in
each spatial dimensionality. This classification has been expected to be
also appliable to interacting systems as long as the strength of the
interactions is sufficiently small as compared to the gap\cite%
{schnyder,kitaev1}. However recent studies on a specific one-dimensional
(1D) model show that the free-fermion classification breaks down in the
presence of interactions\cite{kitaev2,amturner}. So it is most possible that
the interaction doesn't modify the topological nontrivial classes uniformly.

In this paper, we use the ED to study the effect of interactions in a 1D
lattice model which is known to have a topological state in its free form.
With the ground-state energies and wave functions, we firstly calculate the
energy and the distribution of the quasi-particle added or removed from the
system at half-filling for open (OBC) and periodic boundary conditions
(PBC), then identify the topological phases with the existence of the edge
states. The topological features indeed have manifested themselves clearly
in the small sizes we can access. We consider both the repulsive and
attractive interactions and find the topological phase is robust to repulsive interaction
while stabilized by attractive one when the interaction strengths are small. For repulsive interaction we
find that as its strength is increased, the system undergoes a topological
quantum phase transition (TQPT) into a trivial insulator. For attractive interaction it can
drive a trivial insulator into a non-trivial one. Then we calculate the
Berry phase and the parity of the bulk system, which are equivalent in
characterizing the topological property of the system. When the system is in
the topological phase, they have nontrivial values. With them we obtain the
critical interaction strengths and construct part of the phase diagram in the parameters
space. Finally we discuss the effective Hamiltonian at large-$U$ limit and
provide additional understanding of the numerical results.


\section{The 1D model with edge modes}

%
\begin{figure}[tbp]
\includegraphics[width=8cm]{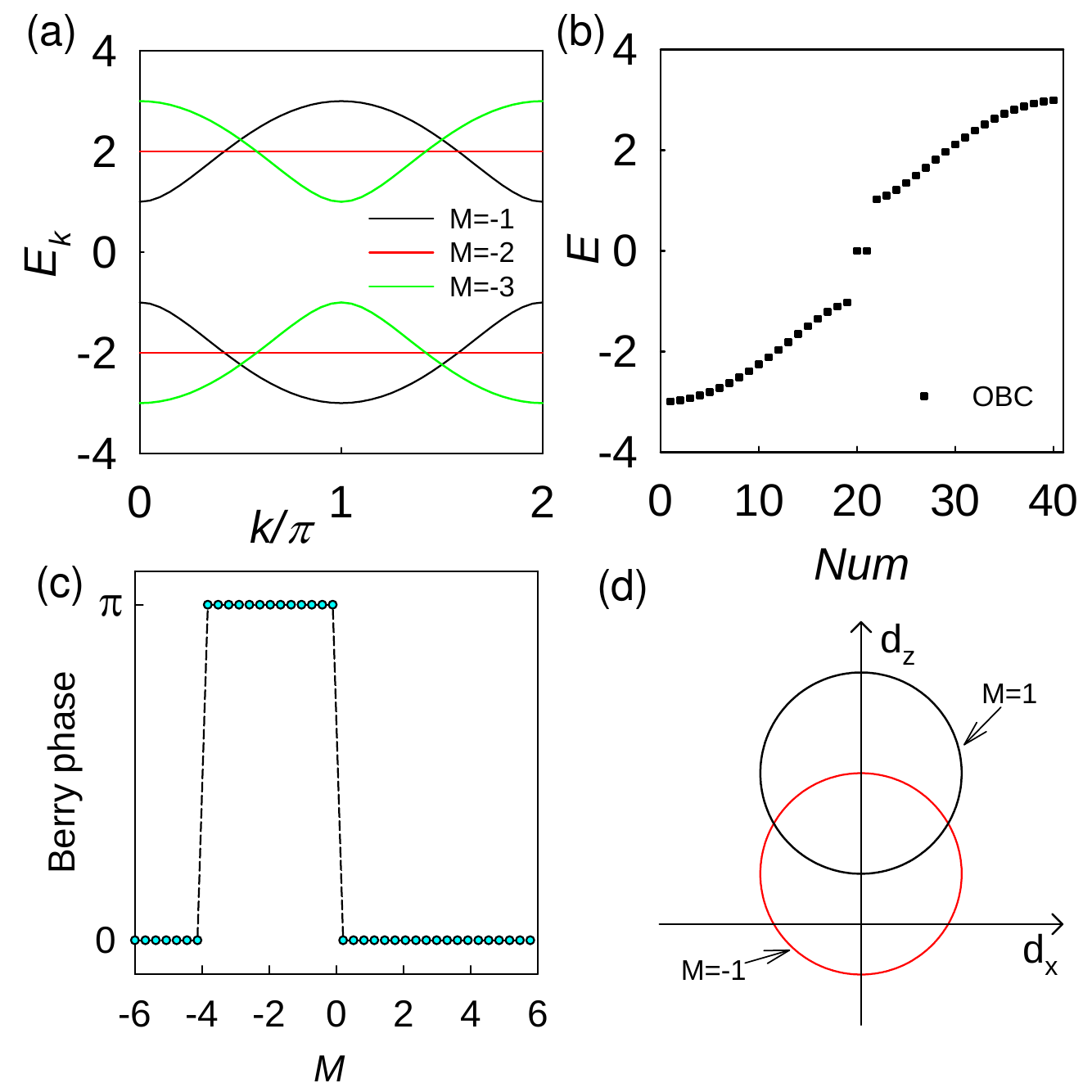}
\caption{(Color online) (a)The tight-binding band structure of ${\cal H}(k)$. The low-energy excitation is located at $k_{1}$ (black line) or $%
k_{2}$ (green line) depending on the parameters. (b) Edge modes in the
topological phase ($M=-1$) on a chain of length $N=20$ with OBC. (c) Berry
phase of the occupied Bloch state of Eq.(\protect\ref{eq1}) at different $M$%
. (d) The plot of the curve $[d_{x}(k),d_{z}(k)]$ with $k\in \lbrack 0,2%
\protect\pi ]$. In all figures $B=1$ is set.}
\label{fig1}
\end{figure}

\begin{figure*}[t]
\begin{minipage}[l]{\linewidth}
    \includegraphics[scale=1]{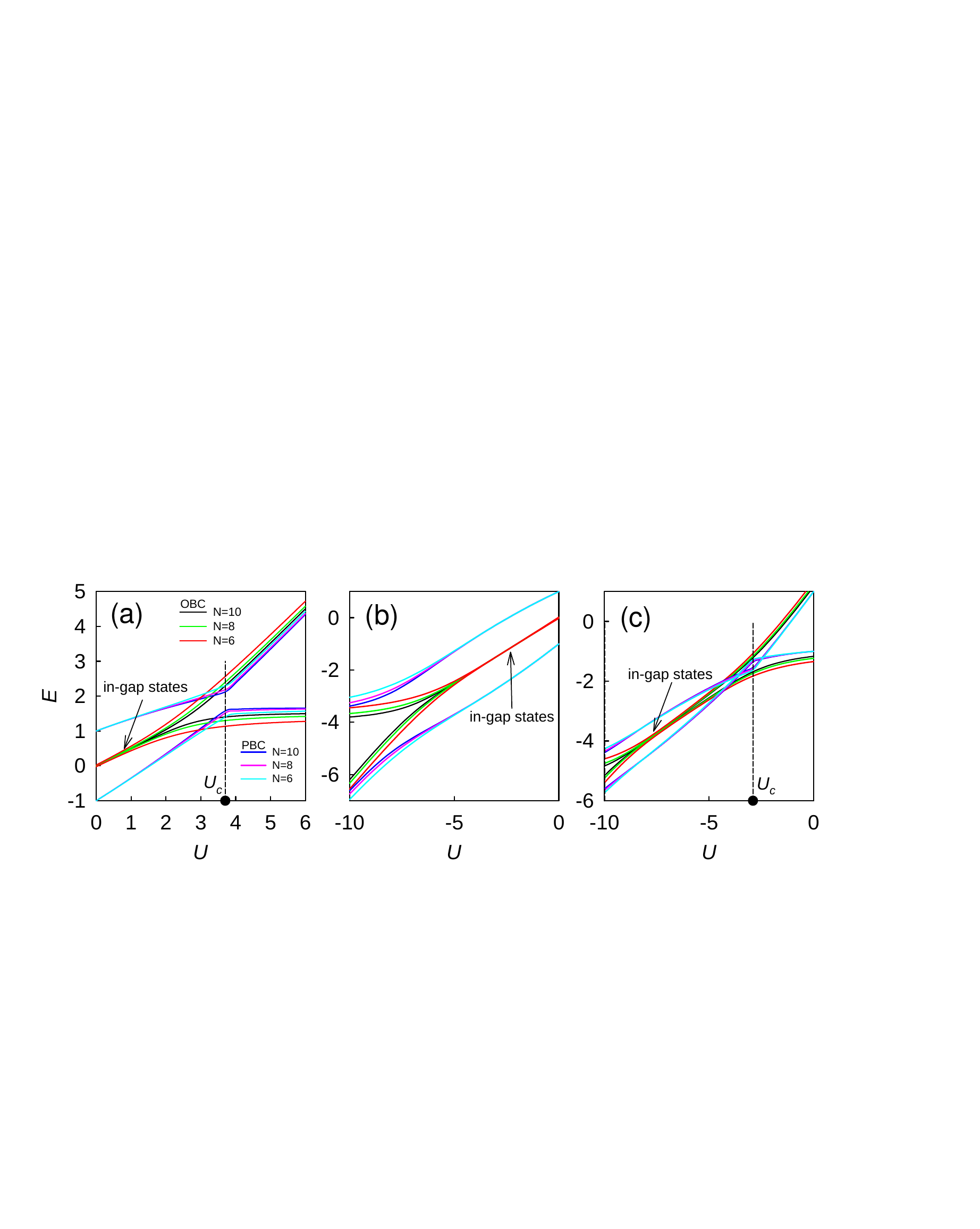}
  \end{minipage}
\vspace{-0pt}
\caption{(Color online) The energies of the quasi-particles added or removed
from the half-filling system with OBC and PBC. (a) and (b): the phase at $%
U=0 $ is topological ($M=-1$); (c): the phase at $U=0$ is trivial ($M=1$).
The critical interactions $U_c$ in (a) and (c) can be accurately determined
from the parity of the ground-state wave function. Here $B=1$.}
\label{fig2}
\end{figure*}

Our starting point is the 1D non-interacting tight-binding model,\cite%
{sqshen1}
\begin{eqnarray}
H_{0} &=&\sum_{i}(M+2B)\Psi _{i}^{\dagger }\sigma _{z}\Psi _{i}-\sum_{i,\hat{%
x}}B\Psi _{i}^{\dagger }\sigma _{z}\Psi _{i+\hat{x}}  \label{eq1} \\
&-&\sum_{i,\hat{x}}sgn(\hat{x})iA\Psi _{i}^{\dagger }\sigma _{x}\Psi _{i+%
\hat{x}}  \nonumber
\end{eqnarray}%
where $\sigma _{x}$, $\sigma _{z}$ are Pauli matrices and $\Psi
_{i}=(c_{i\uparrow },c_{i\downarrow })^{T}$ with $c_{i\uparrow }$($%
c_{i\downarrow }$) electron annihilating operator at the site $\mathbf{r}%
_{i} $. The first two terms represent the differences of the on-site potentials and the hopping amplitudes between the up- and down- electrons, and the third term is due to the spin-orbit coupling. In momentum space Eq.(\ref{eq1}) becomes $H_{0}=\sum_{k}\Psi
_{k}^{\dagger }\mathcal{H}(k)\Psi _{k}$ with $\Psi _{k}=(c_{k\uparrow
},c_{k\downarrow })^{T}$ the Fourier partner of $\Psi _{i}$ and
\[
\mathcal{H}(k)=[M+2B-2Bcos(k)]\sigma _{z}+2Asin(k)\sigma _{x}.
\]%
The spectrum of $\mathcal{H}(k)$ consists of two bands,
\[
E_{k}^{(1,2)}=\pm \sqrt{\lbrack M+2B-2Bcos(k)]^{2}+[2Asin(k)]^{2}}.
\]%
Usually the two bands are dispersive, but when the parameters satisfy $%
-M/2=B=A$ the bands are flat. For $M=0$ ($M=-4B$) bands 1 and 2 touch at the
Dirac point $k_{1}=0$ ($k_{2}=\pi $), while for other values a gap $\Delta $%
=min\{$2|M|$, $2|M+4B|$\} opens up at the Dirac point $k_{1}$ or $k_{2}$. At
half-filling, depending on the values of the parameters $A$, $B$ and $M$ the
system can be a trivial insulator or a non-trivial insulator with edge
modes. In the following of the paper we take $B$ positive and set $A=1$ as
the energy scale.

The topological property of the system can be understood in terms of Berry
phase in $k$ space, which is $\gamma =\oint \mathcal{A}(k)dk$ with the Berry
connection $\mathcal{A}(k)=i\langle u_{k}|\frac{d}{dk}|u_{k}\rangle $ and $%
|u_{k}\rangle $ the occupied Bloch state\cite{resta,dxiao}. The numerical
result is shown in Fig.\ref{fig1}(c). It shows the Berry phase $\gamma $ mod
$2\pi $ gets a nonzero value $\pi $ for $-4B<M<0$. We also have performed
numerical diagonalization of $H_{0}$ on a chain with OBC. In accord with the
above argument, we find a pair of zero modes appearing in the gap when the
Berry phase of the system is $\pi $ (Fig.\ref{fig1}(b)).

We notice that when the Berry phase is $\pi $ the masses at the two Dirac
points $k_{1}$ and $k_{2}$ have different signs, i.e., $M(M+4B)<0$. This can
also serve as a criterion of the topological property in the system. The
reason can be understood from the low-energy Hamiltonians governing the
excitations in the vicinity of the Dirac points\cite{hmguo1,hmguo2,sqshen1,sqshen2}.
By linearizing ${\cal {H}}(k)$ near $k_{1}$ and $k_{2}$ we obtain two Dirac
Hamiltonians,
\begin{eqnarray}
h_{k}^{1} &=&2A\sigma _{x}k+M\sigma _{z}  \label{eq4} \\
h_{k}^{2} &=&-2A\sigma _{x}k+(M+4B)\sigma _{z}.  \nonumber
\end{eqnarray}%
For $-4B<M<0$ the masses at the Dirac points exhibit opposite signs, while
for $M>0$ and $M<-4B$ the masses at the Dirac points exhibit the same signs.
In the following we show that the two cases correspond to two phases with
different topological properties. We consider a junction between the two
phases running along a line in real space (suppose $-4B<M<0$ for $x<0$ and $%
M>0$ for $x>0$). So the mass of $h_{k}^{1}$ necessarily undergoes a sign
change across the $x=0$ boundary. Such a soliton mass profile is known to
produce massless state in the associated Dirac equation, localized near the
boundary. Specifically, Dirac equation
\[
\lbrack 2A(-i)\sigma _{x}\partial _{x}+\sigma _{z}m(x)]\phi (x)=E\phi (x)
\]%
with $m(x\rightarrow -\infty )<0$ and $m(x\rightarrow \infty )>0$ has a
gapless solution
\[
\phi (x)=e^{-\frac{1}{2A}\int_{0}^{x}m(x^{\prime })dx^{\prime }}(%
\begin{array}{c}
1 \\
i%
\end{array}%
)
\]%
localized at the boundary with zero energy. Since the $M>0$ phase can be
continually connected to the $M=\infty $ phase which is a trivial insulator,
the phase for $-4B<M<0$ has non-trivial topological property. The above
argument is similar when $M<-4B$ for $x>0$ in the junction. So through the
relative signs at the Dirac points we can obtain the same condition for the
non-trivial topological phase.

Furthermore, we can write $\mathcal{H}(k)=\mathbf{d}%
(k)\cdot \vec{\sigma}$, where $d_{x}(k)=2A sin(k)$ ,$d_{y}(k)=0$ and $%
d_{z}(k)=M+2B-2B cos(k)$. The pair $[d_{x}(k),d_{z}(k)]$ forms a closed loop
in the plane when $k$ changes from $0$ to $2\pi$. If the system has
non-trivial topological properties, the loop will contain the origin point
of the plane (Fig.\ref{fig1}(d)). It is also consistent with the former
arguments.

\section{The effect of Hubbard interaction}

To study the effect of interaction, we add the Hubbard term $H_{1}=U\sum_{i}
n_{i\uparrow}n_{i\downarrow}$ to Eq.(\ref{eq1}). This term changes the
single electron Hamiltonian Eq.(\ref{eq1}) to a many-body one. For
non-interacting case, the topological property can be directly read from the
presence of the edge states in the gap of the single electron energy
spectrum. However in the presence of interaction, there is no longer single
electron state. Then how can we identify the topological property of the
system?

\subsection{The existence of the edge states}

Generally for a finite chain of $N$ sites, the full Hilbert space of the
system has dimension $4^{N}$. Since in the Hamiltonian $H_{0}+H_{1}$ the
total particle number is conserved, we can get the ground-state energy and wave function of a
system with a fixed number of electrons using ED. Then we can define the
energy of the quasi-particle added to a system with $n$ electrons as $\Delta
E_{n}=E_{n+1}^{0}-E_{n}^{0}$, where $E_{n}^{0}$ is the ground energy of a
system with $n$ particles. Similarly to the non-interacting case, if the
system has non-trivial topological property, there appear states in the gap
of the quasi-particle energy spectrum (QPES) as the boundary condition
changes from PBC to OBC. Since this definition can be continually connected
to the single electron case, we expect it valid at least for small-$U$ cases.

Firstly we consider the effect of repulsive interaction. Figure \ref{fig2}(a) shows the results
of starting from a system with non-trivial topological property. Similar to
the non-interacting case, we concern about the electron added or removed
from the half-filling system. It shows that the energies of the
quasi-particles added or removed appear in the gap when $U$ is below a
critical value $U_c$. At small $U$ the two in-gap modes have exactly the
same values. Due to the finite-size effect, when $U$ approaches the critical
value, the energies of the two modes is separated by a gap. When $U$ is
beyond $U_c$, the in-gap modes disappear and evolve into the bulk ones. The
result clearly shows that the topological phase survives in the presence of
small repulsive interaction and a TQPT is driven by finite repulsive interaction. We also study the case of
starting from a trivial insulator and find that no in-gap modes appear when repulsive interaction is added.

Next we turn to study the effect of attractive interaction. The calculations are straightforward
and the results are shown in Fig.\ref{fig2}(b) and (c). In Fig.\ref{fig2}(b)
attractive interaction is introduced into a system with non-trivial topological property. It
shows that the existence of the in-gap modes persist to quite large strengths. Then
as the strength is further increased, the in-gap modes continuously
evolve into bulk ones. In Fig.\ref{fig2}(c) we show the result of
introducing attractive interaction to a trivial insulator. We find that when the strength
reaches a critical value $U_{c}$, the in-gap modes begin to appear in the
gap of QPES, indicating that attractive interaction can drive a trivial phase into a non-trivial
one. Then the in-gap modes persist till the strength becomes very large
when they evolve into bulk ones. This behavior is very similar to that of
disorder\cite{sqshen,hmguo3}.

\begin{figure}[tbp]
\includegraphics[width=8cm]{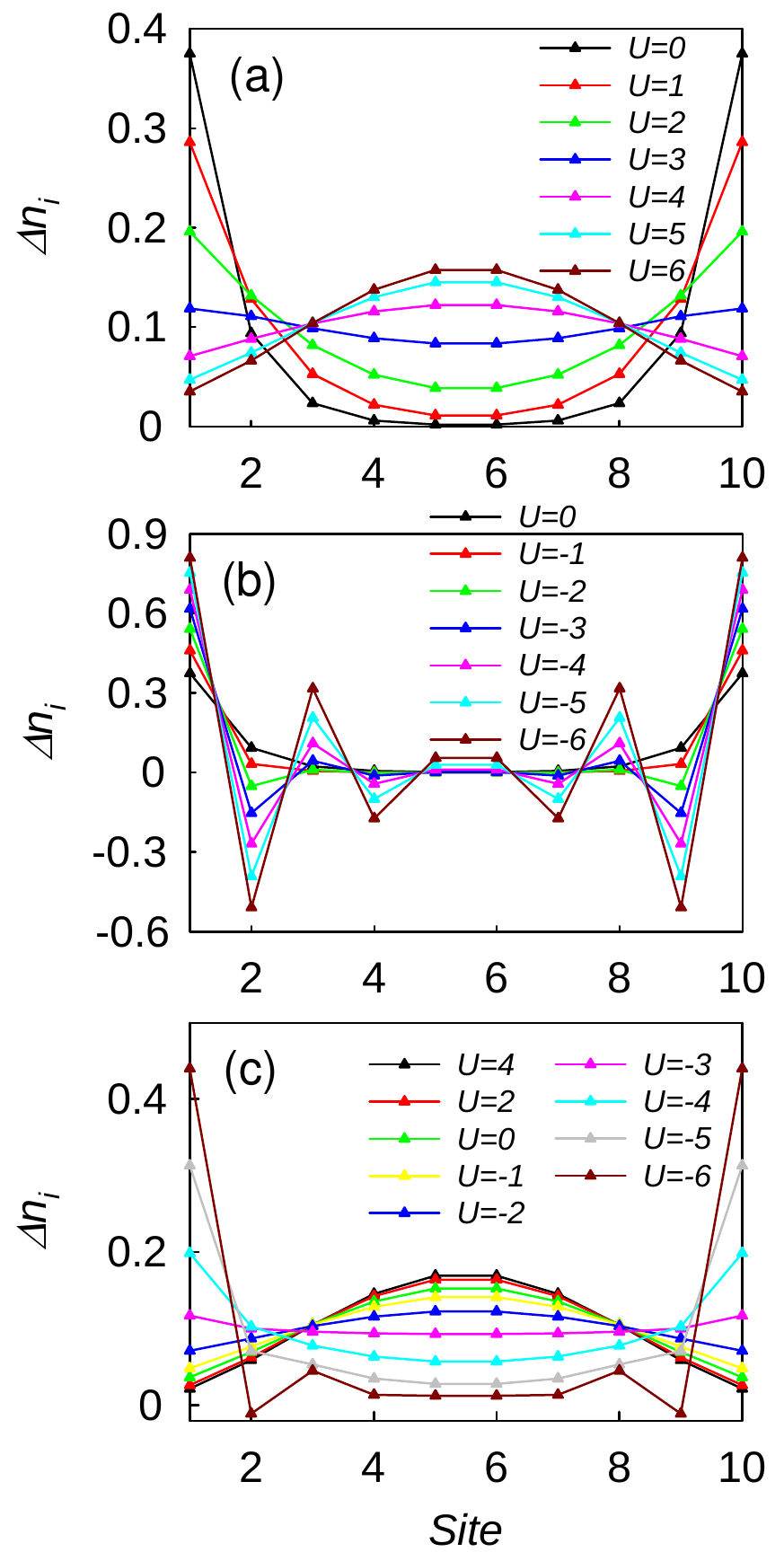}
\caption{(Color online) The distribution of the quasi-particles added or
removed from the half-filling system with OBC. (a) and (b): $M=-1$; (c) $M=1$%
. Here $B=1$ and the system size $N=10$.}
\label{fig3}
\end{figure}

Till now by identifying the in-gap states in the QPES, we show the effects
of repulsive and attractive Hubbard interactions in the topological phase.
In the non-interacting systems, the in-gap mode is also referred to edge
mode due to the fact they mainly distribute near the edges of the chain.
Similarly in the interacting case we can also calculate the distribution of
the in-gap mode to study its nature. Using the many-body wave functions, the
distribution of the electron added can be defined as: $\Delta n_{i}=\langle
\psi^{0}_{n+1}|\hat{n}_{i}|\psi^{0}_{n+1}\rangle-\langle \psi^{0}_{n}|\hat{n}%
_{i}|\psi^{0}_{n}\rangle$, where $\hat{n}_{i}=c^{\dagger}_{i\uparrow}c_{i%
\uparrow}+c^{\dagger}_{i\downarrow}c_{i\downarrow}$ is the electron number
operator on site $i$ and $\psi^{0}_{n}$ is the ground-state wave function of the
system with $n$ electrons. The results at different $U$ and $M$ are shown in
Fig.\ref{fig3}. Since the electrons added or removed at half-filling have
exactly the same distributions, only one of them is shown. In Fig.\ref{fig3}%
(a), we start from a system with nontrivial topological property. As
expected at $U=0$ the in-gap state mainly distributes near the edges. As we
increase the strength of the repulsive interaction, the distribution begins to evolve from the ends to the
bulk. Though the result is greatly affected by the finite-size effect, the
distribution at $U<U_c$ is still clearly distinct from that at $U>U_c$ ($U_c$
is about $3\sim4$), indicating the different topological properties existing
in the system. Next we add attractive interaction to the above system and the result is shown in
Fig.\ref{fig3}(b). As the strength increases, the topological phase is
firstly stabilized, manifested by the increase of the components near the
ends. Then the distribution begins to oscillate between positive and
negative values. The occurrence of negative value is due to the many-body
effect. The amplitude of the oscillation decays from the end to the bulk.
And as the strength is further increased, the decay becomes slower, so
the in-gap state begins to disappear, indicating the system experiencing a
TQPT. We also start from a trivial insulator and the result is shown in Fig.%
\ref{fig3}(c). It shows that in the presence of repulsive interaction the distribution still
mainly concentrates in the bulk, while after attractive interaction is added the distribution
begins to evolve from the bulk to the ends. The dramatic change occurs at $%
U=-3$ when the distribution shows a sign of edge one. The estimated critical
value $-3\sim -2$ is in good consistent with $U_{c}$ in Fig.\ref{fig2}(c).
Then as the strength of the attractive interaction is further increased the distribution shows a similar
behavior as that in Fig.\ref{fig3}(b). Our these results are in good
consistent with those obtained from the QPES.

\begin{figure}[tbp]
\includegraphics[width=8cm]{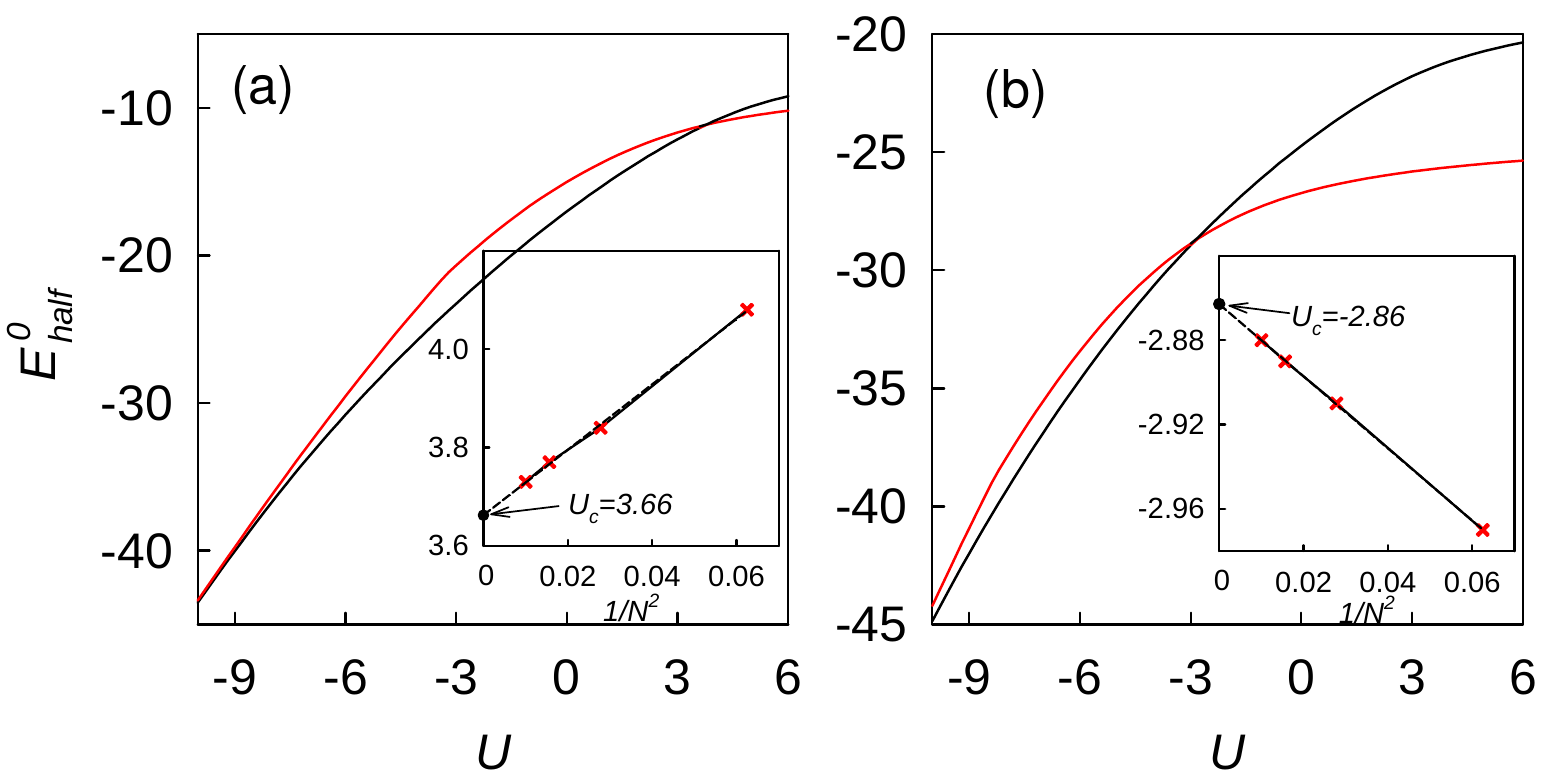}
\caption{(Color online) The energies and parities of the ground- and the
first-excited states. (a)$M=-1$ and (b)$M=1$. The red (black) curve has the
parity value -1(1). In both figures $B=1$ and $N=8$.}
\label{fig4}
\end{figure}

\subsection{The topological properties of the bulk system}

\begin{figure}[tbp]
\includegraphics[width=8cm]{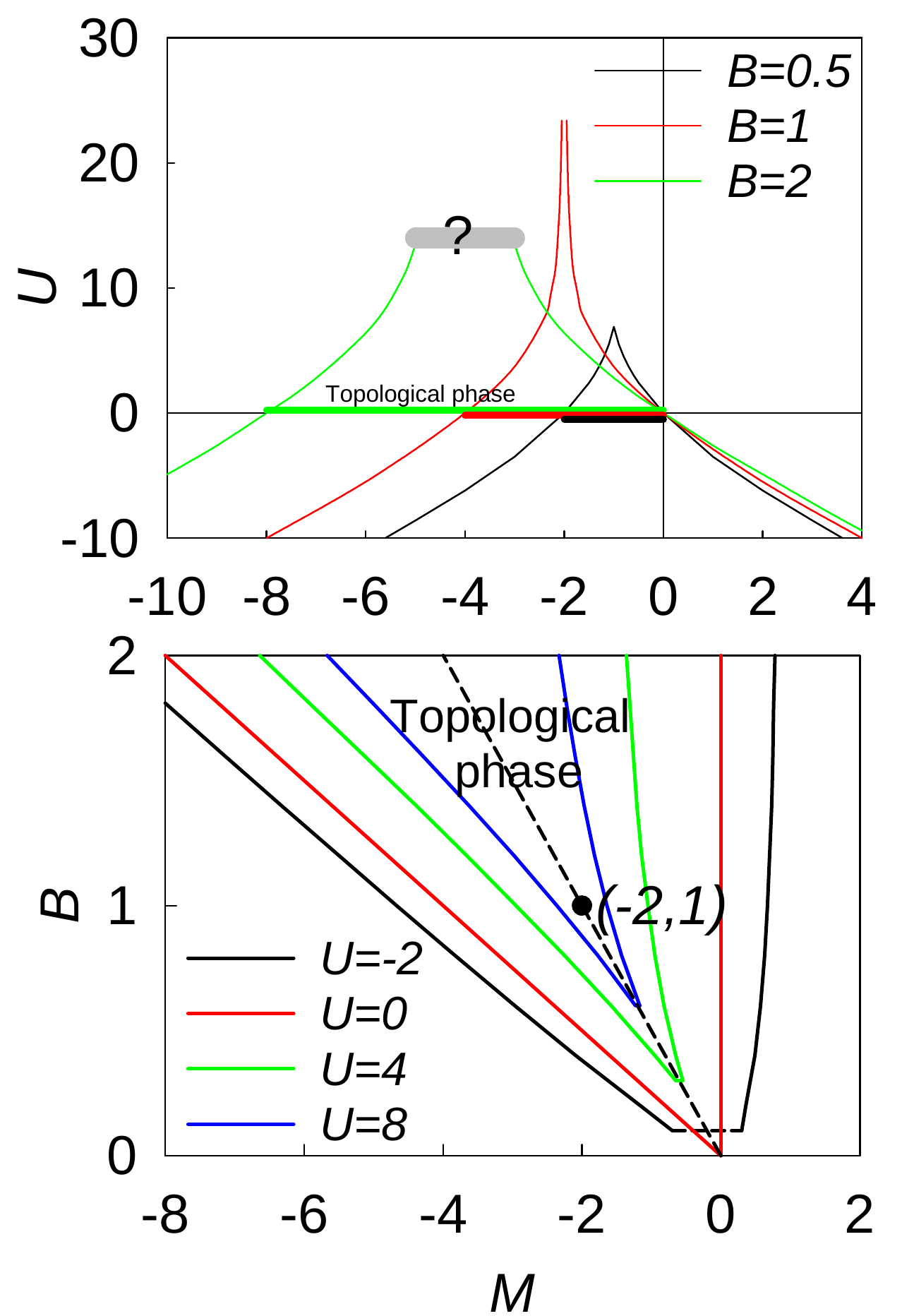}
\caption{(Color online) The phase diagram in $(M,U)$ and $(M,B)$ planes. The
system size $N=8$.}
\label{fig5}
\end{figure}

Till now we have examined the topological phase in the interacting system
with the existence of the edge states. It is known that their existence is
due to the bulk topological properties. So in the following we will
calculate the Berry phase of the ground-state of the interacting system at
half-filling using the twisted boundary conditions\cite{dxiao,qniu,hatsugai}. It can
be defined as
\[
\gamma =\oint i\langle \psi _{\theta }|\frac{d}{d\theta }|\psi _{\theta
}\rangle ,
\]%
where $\theta $ is the twisted boundary phase which takes values from $0$ to
$2\pi $ and $\psi _{\theta }$ is the corresponding ground-state many-body
wave function at half-filling. To compare with the previous results, we
firstly study the Berry phases associated with Fig.\ref{fig2}. We find that
the Berry phase $\gamma $ mod $2\pi $ has the value $\pi $ and $0$ for $%
U<U_{c}$ and $U>U_{c}$ (see Fig.\ref{fig2}).

The Hamiltonian ${\cal H}(k)$ possesses the symmetry
implemented by a unitary transformation
\begin{eqnarray}
\sigma_{z} \mathcal{H}(k) \sigma_{z}=\mathcal{H}(-k).
\end{eqnarray}
The symmetry is similar to the inversion symmetry except that there is an
additional sign when it is performed on the spin-down electron. Thus we can
define the parity of the wave function $\psi^{0}_{n}$. In the basis of fixed
number of electrons, $\psi^{0}_{n}=\sum_{i}\phi_{i}|i\rangle$ with $%
|i\rangle $ denoting the $i$-th $n$-electrons basis. Under the inversion
transformation, $\mathcal{P}\psi^{0}_{n}=\sum_{i}\phi_{i}\mathcal{P}%
|i\rangle=\sum_{i}\phi_{i}\epsilon_{i}|j\rangle_{i}=\pm \psi^{0}_{n}$, where
$\mathcal{P}|i\rangle=\epsilon_{i}|j\rangle_{i}$ and $\epsilon_{i}=\pm 1$
depending on the times of exchanging the fermion operators. We calculate the
parities of the ground- and the first-excited states associated with Fig.\ref%
{fig2} and the results are shown in Fig.\ref{fig4}. It shows that at
half-filling there is a gap between the two states even in the presence of
interactions and the gap may be eliminated at the critical strength $U_{c}$
where the parity also change its sign.

We also find that the parity is the same as the Berry phase to characterize
the bulk topological properties since they both change their values at the
same critical value $U_{c}$. From either one the accurate values of the
critical interactions can be obtained, such as: $U_{c}=3.66$ in Fig.\ref%
{fig2}(a) and $U_{c}=-2.86$ in Fig.\ref{fig2}(c). We want to mention that in
some cases such as at large attractive interaction though the in-gap states disappear, the Berry
phase or the parity of the ground-state doesn't change their values. These cases are beyond our
discussion because the ground-states become degenerate and the systems
aren't insulators any more.

From these calculations, we can get part of the phase diagram in the
parameters space, as shown in Fig.\ref{fig5}. Fig.\ref{fig5}(a) is in $(M,U)$
plane. it shows that at each value of $B$ the curve is symmetric about $%
M=-2B $ and the behaviors are similar as $M$ is far from $-2B$. When $M$
approaches $-2B$, the property is different depending on the value $B$. For $%
B<1$, the critical interaction gets its maximum at $M=-2B$. While for $B>1$
the results become complex and depend greatly on the system sizes, so no
definite conclusions are made. Fig.\ref{fig5}(b) is in $(M,B)$ plane. At $%
U=0 $ the topological phase is in the area restricted by the lines $M=0$ and
$B=-M/4$. Its size is shrunk by repulsive interaction and broadened by attractive one. Corresponding to the
results in Fig.\ref{fig5}(a) at fixed $B$ of each curve the sum of the two
corresponding $M$ is $-4B$ and when $U>0$ and $B<1$ the tips of the curves
fall on the line $B=-M/2$. We notice that $(M,B)=(-2,1)$ is a special point
where the non-interacting system exhibits flat bands. Here at large repulsive interaction the
ground- and the first-excited states also become degenerate and the Berry
phase and parity of the ground-state keep the same value all the way.

\subsection{The effective Hamiltonian at large-$U$ limit}

To understand the phase diagram better, it is helpful to study the system at
large-$U$ limit, when the effective Hamiltonian is\cite{zzli},
\begin{eqnarray}  \label{eq5}
H_{eff}&=&-\sum_{i}
(J_{x}\sigma_{i}^{x}\sigma_{i+1}^{x}+J_{y}\sigma_{i}^{y}%
\sigma_{i+1}^{y}+J_{z}\sigma_{i}^{z}\sigma_{i+1}^{z}) \\
&+&\sum_{i}(M+2B)\sigma_{i}^{z}  \nonumber
\end{eqnarray}
where $J_{x}=(B^2-A^2)/U$,$J_{y}=(B^2+A^2)/U$ and $J_{z}=-J_{x}$. When $A=B$
it is $H_{eff}^1=-J\sum_{i}
\sigma_{i}^{y}\sigma_{i+1}^{y}+Jg\sum_{i}\sigma_{i}^{z}$ with $J=\frac{2B^2}{%
U}$ and $g=\frac{(M+2B)U}{2B^2}$, which is the quantum Ising model\cite%
{sachdev}. By tuning the dimensionless coupling $g$, $H_{eff}^1$ exhibits a
quantum phase transition and the critical point is exactly at $|g|=1$. For $-%
\frac{2B^2}{U}-2B<M<\frac{2B^2}{U}-2B$, the ground state is twofold
degenerate and possess long-range correlations in the magnetic order
parameter $\sigma_{y}$, while beyond the above range of $M$ it is in a
quantum paramagnetic state. This naturally explains the degeneracy at $%
B=-M/2=1$ for large repulsive interaction.


\section{Conclusions}

To conclude, we study the effect of interactions in a 1D topological model
by means of the ED method. The topological features have already manifested
themselves clearly in the small sizes the calculations can access. Our
studies focus on the half-filling system. We examine the topological phases
with the existence of the edge states, which are exhibited from the energy
and distribution of the electron added or removed to a system at
half-filling and with OBC. We show that the topological phase is robust to
small interactions and finite repulsive interaction can drive a topological
non-trivial phase into a trivial one while the attractive interaction can
drive a trivial phase into a non-trivial one.

We calculate the Berry phase and the parity of the ground-state wave
function to study the bulk topological properties. The Berry phase and the
parity have intrinsic connection and are equivalent to describe the
topological properties due to the symmetry in our model. In the cases where the edge state exists, the Berry phase has nontrivial value $\pi $. At
the TQPT points the Berry phase and the parity change their values. From
them we determine the critical interactions and construct part of the phase
diagrams in the parameters space.

Our these results demonstrate the existence of the topological phases in 1D
interacting fermion systems. Though the model we use is artificial, it may
be constructed experimentally using cold atoms trapped in the optical
lattice, which allows one to directly simulate ideal and tunable models. At
present ultracold Fermi gases in a truly 1D regime can be realized using
strong optical lattices and the interactions can be tuned between repulsive
and attractive ones by means of Feshbach resonances\cite{bloch}. With these
developments some of the basic phenomena in the interacting fermion systems
are being studied. Besides, there appear some studies which suggest the
methods to mimics the effect of spin-orbit coupling and produce topological
states of matter in cold-atom systems \cite{goldman,cjwu,sarma}. So it is
very hopeful that our these results are tested in cold-atom experiments.


\section{Acknowledgements}

The authors thank Shiping Feng for helpful discussions. Support for this work came from the Research Grant Council of Hong Kong
under Grant Nos. N\_HKU748/10, HKUST3/CRF/09,  NSFC under Grant
No. 11104189, and the Ministry of Science and
Technology of China under Grant Nos. 2011CBA00102.

\end{document}